\documentclass[twocolumns]{emulateapj}

\usepackage{natbib}
\usepackage{amssymb}
\usepackage{amsmath}
\def\be{\begin{eqnarray}}
\def\ee{\end{eqnarray}}
\usepackage{comment}
\usepackage[dvipsnames]{xcolor}
\usepackage{graphicx,subfigure}
\usepackage{enumerate}
\usepackage{rotating}
\usepackage{longtable}
\usepackage{booktabs}
\usepackage{float}

\usepackage{hyperref}
\hypersetup{colorlinks,urlcolor={blue},linkcolor={blue},citecolor={blue}}

\usepackage[normalem]{ulem}

\shorttitle{VLITE observations of SN\,2004dk}

\shortauthors{A.~Balasubramanian et al.}

\begin{document}

\title{Radio observations of SN\,2004\lowercase{dk} with VLITE confirm late-time re-brightening}
\author{A.~Balasubramanian\altaffilmark{1,*}, A. Corsi\altaffilmark{1}, E. Polisensky\altaffilmark{2}, T. E. Clarke\altaffilmark{2}, N. E. Kassim\altaffilmark{2}}
\altaffiltext{1}{Department of Physics and Astronomy, Texas Tech University, Box 1051, Lubbock, TX 79409-1051, USA}

\altaffiltext{2}{Naval Research Laboratory, Code 7213, 4555 Overlook Avenue, SW, Washington, DC 20375, USA}

\altaffiltext{$^{*}$}{Email : \email{arvind.balasubramanian@ttu.edu}}

\begin{abstract}
\label{abstract}
The study of stripped-envelope core-collapse supernovae (SNe), with evidence for strong interaction of SN ejecta with the circumstellar medium (CSM), provides insights into the pre-supernova progenitor, and a fast-forwarded view of the progenitor mass-loss history. In this context, we present late-time radio observations of SN\,2004dk, a type Ibc supernova located in the galaxy, NGC 6118, at a distance of $d_L \approx 23$\,Mpc. About 10 years after explosion, SN\,2004dk has shown evidence for H$\alpha$ emission, possibly linked to the SN ejecta interacting with an H-rich CSM. Using data from the VLA Low Band Ionosphere and Transient Experiment (VLITE), we confirm the presence of a late-time radio re-brightening accompanying the observed H$\alpha$ emission. We model the SN\,2004dk radio light curves within the (spherically symmetric) synchrotron-self-absorption (SSA) model. Within this model, our VLITE observations combined with previously collected VLA data favor an interpretation of SN\,2004dk as a strongly CSM-interacting radio SN going through a complex environment shaped by non-steady mass-loss from the SN progenitor. 
\end{abstract}


\section{Introduction}
\label{intro}
Type Ibc supernovae (SNe) are a class of massive star core-collapse explosions that lack hydrogen (H) features in their spectra and are thus referred to as stripped-envelope core-collapse SNe. They  account for $\approx 16.3 \%$ of the local core-collapse SN population \citep[see Figure 6 in][]{2020ApJ...904...35P}. The presence or absence of helium (He) features allows for further classification of these core-collapse explosions into SNe of type Ib (He-rich) and type Ic (He-poor)  \citep[see e.g.,][]{1997ARA&A..35..309F}.
The physical processes that drive mass loss in evolved massive stars that constitute the pre-SN progenitors, as well as the physics of the pre-supernova progenitors themselves, are yet to be understood. Several models have been proposed which include ejection of envelopes of massive Wolf-Rayet (WR) stars through strong dense winds \citep{1986ApJ...302L..59B,1995ApJ...448..315W}, or accretion from a nearby companion which strips the massive star of its H-rich layer \citep{1985ApJ...294L..17W,1992ApJ...391..246P}. 

Studying the properties of the circumstellar medium (CSM) of Ibc SNe is crucial to shed light on the physics of mass loss, and can provide valuable clues on the pre-SN progenitor. 
Following the collapse of the core, a  shock wave travels outwards and reaches the outer envelopes of the star. The emission in X-rays (as well as UV and light at blue visible wavelengths) that accompanies the so-called shock breakout, can help constrain the progenitor radius  \citep{2013ApJ...769...67P,2019ApJ...887..169H}.  Subsequently, the shock front begins interacting with matter ejected by the pre-SN progenitor. Radio (and X-ray) observations can be used to probe this interaction,  providing information on the mass loss history of the stellar progenitor before its SN explosion \citep[e.g.,][]{2005ApJ...621..908S,2006ApJ...651.1005S,2014ApJ...782...42C,2017ApJ...835..140M,2018MNRAS.480L.146R,2019ApJ...872..201P}.

In this paper we report on the late-time radio observations of SN\,2004dk, a Type Ib SN characterized by radio emission with a late-time re-brightening. About 10 years  after  explosion,  SN\,2004dk  has  shown  evidence  of  H$\alpha$ emission likely linked to the SN ejecta interacting with an H-rich CSM \citep{2019ApJ...883..120P,2018MNRAS.478.5050M}.  X-ray emission associated with SN\,2004dk was detected between about 10 days and 14.5 years since explosion \citep{2007AIPC..937..381P,2019ApJ...883..120P}. Radio monitoring at GHz frequencies up to about 4.5 years since explosion revealed a GHz flux that initially decayed with time, followed by a re-brightening \citep[see][]{2009CBET.1714....1S,2012ApJ...752...17W}. The early-time radio light curve evolution agreed with a synchrotron self-absorbed (SSA) model where the SN shock propagates into a steady-state wind with a density profile scaling as $
\rho \propto r^{-2}$ \citep{2006ApJ...651..381C}. The radio re-brightening observed around 4.5\,yrs since explosion was interpreted also within the SSA model, and attributed to the collision of the forward shock with a denser CSM \citep[see][]{2009CBET.1714....1S,2012ApJ...752...17W}. More recently,  \citet{2019ApJ...883..120P} have analyzed the late-time optical and X-ray observations of SN\,2004dk in the ``wind bubble model'', where the late-time emission results from the SN shock interaction with a complex CSM that has been previously structured by interacting fast and slow winds.

Here, we present new radio observations of SN\,2004dk carried out about 14 years after explosion with the VLA Low Band Ionosphere and Transient Experiment  (VLITE\footnotemark[3]; \citealt{2016SPIE.9906E..5BC}). \footnotetext[3]{\url{https://vlite.nrao.edu/}}SN\,2004dk is the first SN of its type to be detected with VLITE. Low-frequency radio observations are critical to constrain the absorption mechanisms and the CSM structure. We thus re-analyze the radio dataset of SN\,2004dk within the SSA model including our low-frequency observations.

Our paper is organized as follows. In Section \ref{sec:obs}, we summarize previous observations of SN\,2004dk that are relevant for this study, present our VLITE observations and briefly describe the data reduction process. In Section \ref{sec:model}, we describe the SSA model used for our analysis. In Section \ref{sec:results}, we present the results of our light curve fits for SN\,2004dk. In Section \ref{sec:disc}, we discuss our results and conclude.

%%%%%%%%%%%%%%%%%%%%%%%%%%%%%%%%%%%%%%%%%%%%%%%%%%%%%%%%%%%%%%%%%%%%%%%%%%%%%%%%%%

\begin{figure*}\label{fig:radio_contours}
  \begin{center}
  \centering
  \leavevmode
  \hbox{
  \includegraphics[width=0.35\textwidth]{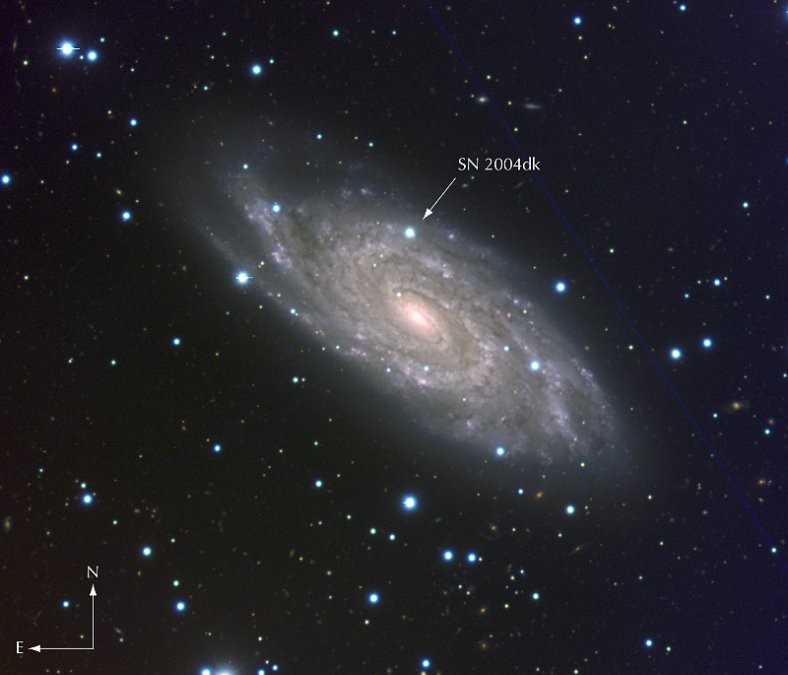}
  \includegraphics[width=0.3\textwidth]{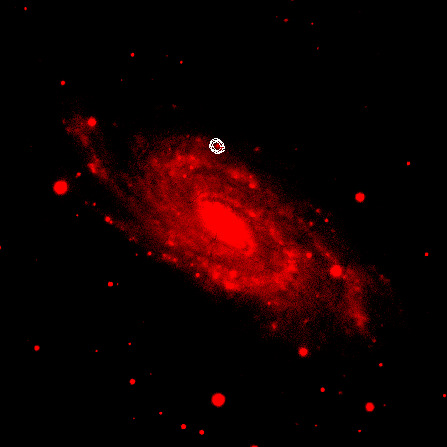}
  \includegraphics[width=0.3\textwidth]{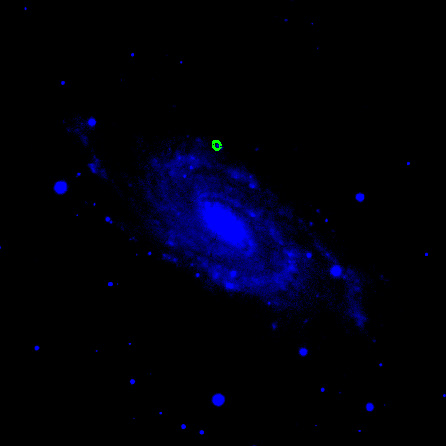}
  }
\caption{\emph{Left}: VLT MELIPAL$+$VIMOS image (\url{www.eso.org/public/images/eso0436b/}) of the spiral galaxy NGC\,6118 and SN\,2004dk taken on 21 August 2004. \emph{Center}: Radio contours observed using VLITE (white - 10$\sigma$ outermost contour, see Table \ref{tab:data} for the data) on 01 Jan 2021\,UT and  \emph{Right}: VLASS (green - 7$\sigma$ outermost contour, see Table \ref{tab:data} for the data) on 21 May 2019, both overplotted on an r-band Pan-STARRS1 image of the field \citep{2020ApJS..251....7F} with the help of SAOImageDS9 \citep{2003ASPC..295..489J}.}%
\end{center}
\vspace{0.59cm}
\end{figure*}

\section{Observations and data reduction}\label{sec:obs}

SN\,2004dk was discovered by the the Katzman Automatic Imaging Telescope (KAIT) on 2004 August 1.19\,UT (thus, hereafter we assume an explosion time of $t_{e}$= 2453216.69\,JD (J2000), 2\,days before August 1.19 \citep[see][]{2004CBET...75....1G} at RA=16h21m48.87s, Dec=-02d16m17.6s . SN\,2004dk is located in the galaxy, \object{NGC 6118}, at a distance of $\approx$ 23\,Mpc \citep{2012ApJ...752...17W}. Subsequent spectroscopic observations helped classify this SN as a He-rich, H-poor, Type Ib \citep{2004IAUC.8404....1F}. SN\,2004dk was also detected in X-rays starting from about 10 days since optical discovery \citep{2007AIPC..937..381P}. Signs of CSM interaction in SN\,2004dk were reported first at wavelengths other than optical. The SN radio luminosity was observed to have re-brightened at $\approx 1700$\,d since explosion \citep{2009CBET.1714....1S} and to persist to at least $\approx 1900$\,d \citep{2012ApJ...752...17W}. Subsequently, late-time optical observations led to the detection of a narrow-band H$\alpha$ excess from SN\,2004dk \citep{2017ApJ...837...62V}. \citet{2018MNRAS.478.5050M} then identified broad H$\alpha$ emission, suggesting that the re-brightening observed at $\sim 1700$\,days can be attributed to the interaction of the SN ejecta with the previously ejected H-envelope by the progenitor. In what follows, we discuss in more detail the observation of SN\,2004dk at radio wavelengths, and present our new data collected with VLITE.

\subsection{VLA observations}
SN\,2004dk was first observed at GHz radio frequencies at about 8\,days after explosion, with the Karl G. Jansky Very Large Array (VLA). Follow-up observations with the VLA continued until  2009  \citep[$\Delta t \sim1912$\,days since explosion; ][]{2009CBET.1714....1S,2012ApJ...752...17W}. A summary of the radio observations carried out with the VLA between 4.9\,GHz and 22.5\,GHz is reported in the first section of Table \ref{tab:data} \citep[adapted from][]{2012ApJ...752...17W}. The 8.5\,GHz light curve of SN\,2004dk peaks at $\sim$ 14\,days from the SN explosion with a flux density $F_{\nu} \approx$ 2.5\,mJy. At an epoch of $\Delta t \sim1700$\,days since explosion, a re-brightening is observed at 4.9\,GHz and 8.5\,GHz. 

A more recent GHz radio observation of the SN\,2004dk field was carried out on 2018 April 7.4 UT ($\Delta t \sim 5000$\,days since explosion) with the VLA in its A configuration (VLA/18A-119, PI: Margutti; see Table\,\ref{tab:data}). 3C286 was used as flux calibrator and J1557-0001 was used as complex gain calibrator. We calibrated this data in \texttt{CASA} \citep{McMullin2007} using the standard VLA calibration pipeline, and then manually inspected and flagged the calibrated data for  Radio Frequency Interference (RFI). After imaging the field with the \texttt{clean} task interactively, we used the \texttt{CASA} task \texttt{imfit} to obtain the integrated flux density of SN\,2004dk (and the associated statistical error) within a circular region of radius 0.33\,\arcsec (nominal VLA resolution at the observed frequency for A-array configuration) centered on the optical position of the SN. A 5\% absolute flux calibration error was added in quadrature to the flux density error returned by \texttt{imfit} \citep[e.g.,][]{2011ApJ...740...65O}. 

\subsection{VLITE observations}

VLITE is a commensal, low-frequency system on the VLA that runs in
parallel (since 2014) with nearly all observations above 1\,GHz. VLITE provides real-time correlation of
the signal from a subset of VLA antennas using the low-band receiver system \citep{2011ursi.confE...5C} and a dedicated DiFX-based software correlator \citep{2007PASP..119..318D}. The VLITE system processes 64\,MHz of
bandwidth centered on 352\,MHz, but due to strong RFI in the upper portion of the band, the usable frequency range is limited to an RFI-free band of $\sim$40\,MHz, centered on 338\,MHz.

SN\,2004dk was observed by VLITE on 2018 April 8, 2019 August 30, 2020 December 28 and 2021 January 1 UT. These dates correspond to $\Delta t\sim$ 5000, 5500 and 6000 days after the
SN\,2004dk explosion, respectively. The VLITE data are reported in the bottom section of Table \ref{tab:data}. SN\,2004dk is the first SN to be detected at sub-GHz frequencies by VLITE.  VLITE data were processed using a dedicated calibration pipeline based on a combination of Obit \citep{2008PASP..120..439C} and AIPS \citep{1996ASPC..101...37V} data reduction packages. The calibration pipeline uses standard automated tasks for the removal of RFI and follows common techniques of radio-interferometric data reduction, including delay, gain, and bandpass calibration (for details on the pipeline data reduction see \cite{2016ApJ...832...60P}). The flux density scale is set using \cite{2017ApJS..230....7P}. The final VLITE images have restoring beams of approximately 6$^{\prime\prime}\times$4$^{\prime\prime}$ and rms noise $0.5-1$\,mJy beam$^{-1}$ (1$\sigma$). Flux densities were measured with PyBDSF \citep{2015ascl.soft02007M} via the VLITE Database Pipeline \citep{2019ASPC..523..441P}. We conservatively adopt 20\% flux density errors that include local image noise as well as fitting and flux-scale uncertainties. 

\subsection{VLASS archival observation}
The VLA Sky Survey (VLASS) is an ongoing $2-4$\,GHz survey of the entire sky visible to the VLA i.e., $\delta > -40$\,deg for a total of about 33,885\,deg$^{2}$ of the sky \citep{2020PASP..132c5001L}. 

We queried the VLASS quick look image archive for fields centered at $<1$\,deg from the optical position of SN\,2004dk. We manually inspected the retrieved VLASS images and identified a source coincident with the optical position of SN\,2004dk. The VLASS field containing the SN (observed in VLASS epoch 1.2\footnotemark[4]) was then analyzed using the \texttt{CASA} task \texttt{imfit} to obtain the integrated source flux density (and corresponding statistical error), using a circular region of radius 2.5\,\arcsec (nominal VLASS resolution) centered at the optical SN position. In order to account for absolute flux calibration errors, we added in quadrature to the integrated flux error returned by \texttt{imfit} a fractional 10\% error (as VLASS Epoch 1 total flux densities are estimated to have systematic errors of order 10\%\footnotemark[4]).
\interfootnotelinepenalty=100
\footnotetext[4]{\url{https://science.nrao.edu/science/surveys/vlass/vlass-epoch-1-quick-look-users-guide}}
Our results are reported in the bottom section of Table \ref{tab:data}, and Figure \ref{fig:radio_contours} (right panel) shows the VLASS radio contours of SN\,2004dk.

\renewcommand\arraystretch{1.5}
\setlength\LTcapwidth{2\linewidth}
\begin{longtable*}{llcccccccl}
\caption{Measured radio flux densities of SN\,2004dk until Jan 2021. Data until Oct 2009 UT are from \cite{2012ApJ...752...17W}. The last four rows contain the more recent VLITE and VLA observations presented for the first time in this paper.\label{tab:data}}\\
\toprule
\toprule
Date & Epoch & 340\,MHz & 3.0\,GHz & 4.9\,GHz & 6.0\,GHz & 8.5\,GHz & 15.0\,GHz & 22.5\,GHz & Inst. \\
(UT) & (days) & (mJy) & (mJy) & (mJy) & (mJy) & (mJy) & (mJy) & (mJy) & \\
\midrule
2004 Aug 7.1 & 7.9 & ... & ... & ... & ... & 1.30$\pm$0.05 & ... & ... & VLA-D\\
2004 Aug 11.0 & 11.8 & ... & ... & 0.87$\pm$0.06 & ... & 2.11$\pm$0.06 & ... & 2.78$\pm$0.33 & VLA-D\\
2004 Aug 13.0 & 13.8 & ... & ... & 1.09$\pm$0.06 & ... & 2.42$\pm$0.06 & 3.57$\pm$0.26 & 2.72$\pm$0.37 & VLA-D\\
2004 Sep 2.0 & 33.8 & ... & ... & 1.80$\pm$0.09 & ... & 1.56$\pm$0.13 & 0.79$\pm$0.21 & 0.69$\pm$0.18 & VLA-D\\
2004 Sep 10.0 & 41.8 & ... & ... & 1.06$\pm$0.07 & ... & 0.81$\pm$0.15 & ... & $\leq$0.43 & VLA-A\\
2004 Sep 18.0 & 49.8 & ... & ... & 0.86$\pm$0.10 & ... & 1.19$\pm$0.11 & ... & $\leq$0.47 & VLA-A\\
2004 Oct 4.1 & 65.9 & ... & ... & 0.46$\pm$0.15 & ... & 0.41$\pm$0.09 & $\leq$0.34 & $\leq$1.41 & VLA-A\\
2004 Oct 17.8 & 79.6 & ... & ... & 0.56$\pm$0.10 & ... & 0.47$\pm$0.12 & $\leq$0.92 & ... & VLA-A\\
2005 Feb 12.4 & 197.2 & ... & ... & $<$0.26 & ... & $<$0.17 & ... & ... & VLA-BnA \\
2009 Feb 12.6 & 1658.4 & ... & ... & 0.22$\pm$0.06 & ... & ... & ... & ... & VLA-B\\
2009 Feb 24.5 & 1670.3 & ... & ... & 0.21$\pm$0.03 & ... & 0.19$\pm$0.03 & ... & ... & VLA-B\\
2009 Apr 2.6 & 1707.4 & ... & ... & 0.34$\pm$0.07 & ... & $\leq$0.08 & ... & ... & VLA-B\\
2009 Apr 19.0 & 1723.8 & ... & ... & 0.21$\pm$0.02 & ... & 0.15$\pm$0.02 & ... & ... & VLA-C\\
2009 Sep 19.8 & 1877.6 & ... & ... & 0.28$\pm$0.05 & ... & 0.17$\pm$0.02 & ... & $\leq$0.57 & VLA-B\\
2009 Oct 26.8 & 1914.6 & ... & ... & ... & ... & ... & ... & $<$0.27 & VLA-D \\
\midrule
2018 Apr 8.4 & 5000.2 & 17.10$\pm$3.42 & ... & ... & ... & ... & ... & ... & VLITE\\
2018 Apr 8.4 & 5000.2 & ... & ... & ... & 4.06$\pm$0.24 $^{a}$ & ... & ... & ... & VLA-A\\
2019 May 21.4 & 5408.2 & ... & 6.46$\pm$0.65 $^{b}$ & ... & ... & ... & ... & ... & VLASS\\
2019 Aug 31.0 & 5509.8 & 11.20$\pm$2.24 & ... & ... & ... & ... & ... & ... & VLITE\\
2020 Dec 28.7 & 5995.5 & 21.53$\pm$4.31 & ... & ... & ... & ... & ... & ... & VLITE\\
2021 Jan 1.8 & 5999.6 & 21.52$\pm$4.30 & ... & ... & ... & ... & ... & ... & VLITE\\
\bottomrule
\multicolumn{10}{l}{$^{a}$ Includes additional 5\% flux error as the observation was dynamic range limited.}\\
\multicolumn{10}{l}{$^{b}$ Includes additional 10\% flux error as this is a quick-look observation from VLASS.}
\end{longtable*}

%%%%%%%%%%%%%%%%%%%%%%%%%%%%%%%%%%%%%%%%%%%%%%%%%%%%%%%%%%%%%%%%%%%%%%%%%%%%%%%%%%

\section{Radio light curve modelling}\label{sec:model}
The SSA model \citep{1998ApJ...499..810C,2006ApJ...651..381C} has been used successfully to describe late-time radio emission powered by SN ejecta-CSM interaction \citep[e.g.,][]{2005ApJ...621..908S,2012ApJ...752...17W, 2019ApJ...872..201P}. 
Within the SSA model, for a smooth CSM profile such as the one predicted to arise from a constant mass-loss rate, constant velocity pre-SN wind, the observed radio emission is characterized by a smooth  turn-on  first  at  higher  frequencies,  and  later  at  lower frequencies. This evolution can be explained as a consequence of a self-absorption frequency which decreases with time as the SN shock propagates toward lower density regions. Modifications to this scenario include an initial exponential rise of the low-frequency radio flux due to free–free absorption in the ionized CSM \citep[an effect usually relevant at early times; see e.g.][]{1990ApJ...364..611W,1998ApJ...499..810C}, as well as flux variations associated with a non-smooth CSM profile, perhaps associated with eruptive mass-loss from the SN progenitor \citep{2012ApJ...752...17W,2014ApJ...782...42C, 2019ApJ...872..201P}. 
A detailed description of the SSA model focused on its  application to radio SNe can be found in \cite{2005ApJ...621..908S}. In what follows, we  summarize briefly the model so as to introduce notation relevant for our analysis of the SN\,2004dk data.

Within the SSA model, we consider emission produced at a certain epoch $t$ since explosion (i.e., $t=0$ at the time of the explosion of the SN, $t_e$) from a thin shell of radiating electrons of radius $r$, and thickness $r/\eta$. The electrons are accelerated to a power-law distribution of Lorentz factors $N(\gamma) \propto \gamma^{-p}$, above a minimum Lorentz factor $\gamma_{m}$. The temporal evolution of the radius of the shell, of the magnetic field, and of the minimum Lorentz factor are assumed to follow the relations:
\begin{equation}
    r(t) = r_{0} \Bigg( \frac{t}{t_{0}}\Bigg)^{\alpha_{r}},
\end{equation}
\begin{equation}
    B(t) = B_{0} \Bigg( \frac{t}{t_{0}}\Bigg)^{\alpha_{B}},
\end{equation}
\begin{equation}
    \gamma_{m}(t) = \gamma_{m,0} \Bigg( \frac{t}{t_{0}}\Bigg)^{\alpha_{\gamma}},
\end{equation}
where $t_{0}$ is an arbitrary reference epoch, $t$ is the epoch since explosion $t_e$, and $r_{0}$, $B_{0}$, $\gamma_{m,0}$ are the values of the shell radius, magnetic field, and electron minimum Lorentz factor at this reference epoch, respectively. Also, the density profile of the radiating electrons within the shocked CSM is assumed to be $n_{e} \propto r^{-s}$. Assuming also equipartition, i.e. $\epsilon_{e}=\epsilon_{B}$ for the fraction of energy going into relativistic electrons and magnetic field respectively \citep[see e.g.,][]{2012ApJ...752...17W}, one has:
\begin{equation}
    \alpha_{B} = \frac{(2-s)}{2}\alpha_{r} - 1 \hspace{0.5cm} \text{and} \hspace{0.5cm} \alpha_{\gamma} = 2(\alpha_{r} - 1).
\end{equation}

The flux density at a certain epoch $t$ and frequency $\nu$ is given by:
\begin{equation}
    f_{\nu} = C_{f} \Bigg( \frac{t}{t_{0}}\Bigg)^{(4\alpha_{r}-\alpha_{B})/2} \Big[ 1 - e^{-\tau_{\nu}^{\xi}(t)}\Big]^{1/\xi} \nu^{5/2} F_{3}(x) F_{2}^{-1}(x),
    \label{eq:fnu_original}
\end{equation}
where the optical depth $\tau_{\nu}(t)$ reads:
\begin{equation}
    \tau_{\nu} = C_{\tau} \Bigg( \frac{t}{t_{0}}\Bigg)^{(p-2)\alpha_{\gamma} + (3+p/2)\alpha_{B} + \alpha_{r}} \nu^{-(p+4)/2} F_{2}(x),
\end{equation}
and where $\xi \in [0,1]$ controls the sharpness of the spectral break between optically thin and thick regimes. In the above equations, $F_{2}$ and $F_{3}$ are Bessel functions of $x=2/3(\nu/\nu_{m})$ with $\nu_{m}$ the critical synchrotron frequency of electrons with Lorentz factor equal to $\gamma_m$. This frequency, $\nu_{m}$, evolves with time as:
\begin{equation}
    \nu_{m} = \nu_{m,0} \Bigg( \frac{t}{t_{0}}\Bigg)^{(10\alpha_{r} - s\alpha_{r} - 10)/2},
\end{equation}
where
\begin{equation}
    \nu_{m,0}=\gamma^2_{m,0}\frac{eB_0}{2\pi m_e c}.
\end{equation}
The normalization constants for the flux density and optical depth, $C_{f}$ and $C_{\tau}$, are themselves functions of  $B_0$, $\nu_{m,0}$, $r_0$ \citep[see Equations (A13)-(A14) in][]{2005ApJ...621..908S}:
\begin{equation}
    C_{f} = \frac{2\pi m_{e}}{2+p}  \Bigg( \frac{r_{0}}{d}\Bigg)^{2} \Bigg(\frac{2\pi m_{e}c}{eB_{0}}\Bigg)^{1/2},
\end{equation}
\begin{equation}
    C_{\tau} = \frac{(p+2)(p-2)\gamma_{m,0}^{(p-1)}}{4\pi \eta}  \frac{B_{0}^{2}}{8\pi} \frac{e^{3}B_{0}r_{0}}{m_{e}^{3}c^{4}\gamma_{m,0}} \Bigg( \frac{eB_{0}}{2\pi m_{e} c}\Bigg)^{p/2},
\end{equation}
where $d$ is the distance to the source.

At high frequencies, the observed spectral flux density needs to be corrected for the effects of synchrotron cooling, which occurs at frequencies above the cooling frequency ($\nu_{c}$), defined as: 
\begin{equation}
    \nu_{c} = \frac{18\pi m_{e} c e}{t^{2}\sigma_{T}^{2}B^{3}}.
\end{equation}

Overall, the resulting spectral shape and temporal evolution of the flux density from Equation (\ref{eq:fnu_original}), corrected for the effects of synchrotron cooling, can be approximated as:
\begin{equation}\label{eq:f_nu}
    f_{\nu} \propto
    \begin{cases}
    \nu^{2} t^{2\alpha_{r}+\alpha_{\gamma}} & \nu < \nu_{m},\\
    \nu^{1/3} t^{(4\alpha_{r} - \alpha_{B})/2} & \nu_{m} < \nu < \nu_{a},\\
    \nu^{-(p-1)/2} t^{[4\alpha_{r} - \alpha_{B} + (4+p)\alpha_{\nu_{a}}]/2} & \nu_{a} < \nu < \nu_{c},\\
    \nu^{-p/2} t^{[2\alpha_{r}+(1-3p)\alpha_{B} + (2+p/2)\alpha_{\nu_{a}} - 2p + 1]} & \nu > \nu_{c}\\ 
    \end{cases}
\end{equation}
where $\nu_{m} << \nu_{a}$ (self-absorption frequency), typically a good assumption at late times, and: \begin{eqnarray}\label{eq:nu_a}
   \nonumber  \nu_{a} = \nu_{a,0}\Bigg(\frac{t}{t_{0}}\Bigg)^{[2(p-2)\alpha_{\gamma} + 2(3+p/2)\alpha_{B} + 2\alpha_{r}]/(p+4)}\\=\nu_{a,0}\left(\frac{t}{t_0}\right)^{\alpha_{\nu_a}}.
\end{eqnarray} 
Note that in the above Equation, $\nu_{a,0}$ can be calculated using the condition $\tau_{\nu_{a,0}}(t_0) = 1$.

Based on the above considerations, for a given $t_e$, via comparison with flux density observations at different times and frequencies, one can determine the set of parameters ($B_{0}$, $\alpha_{r}$, $p$, $\nu_{m,0}$, $r_{0}$, $s$, $\xi$). These, in turn, constrain the physics of the emitting shell and the properties of the CSM. Specifically, one can derive constraints \citep[see e.g.,][]{2006ApJ...651.1005S} on the ejecta energy,
\begin{equation}
    E = \frac{4\pi}{\eta} r_{0}^{3} \frac{B_{0}^{2}}{8\pi \epsilon_{e}} \Bigg( \frac{t}{t_{0}}\Bigg)^{(5\alpha_{r} - s\alpha_{r} - 2)}
\end{equation}
electron density, 
\begin{equation}\label{eq:n_e}
    n_{e} = \frac{(p-2)}{(p-1)} \frac{B_{0}^{2}}{8\pi m_{e} c^{2} \gamma_{m,0}} \Bigg( \frac{r}{r_{0}}\Bigg) ^ {-s},
\end{equation}
and the mass loss rate,
\begin{equation}\label{eq:M_dot}
    \Dot{M} = \frac{8\pi n_{e,0} m_{p} r_{0}^{2}v_{w}}{\eta} \frac{(p-2)}{(p-1)} \frac{B_{0}^{2}}{8\pi m_{e} c^{2} \gamma_{m,0}} \Bigg( \frac{r}{r_{0}}\Bigg)^{(2 - s)}.
\end{equation}
In the above Equation, $v_{w}$ is the wind velocity and $n_{e,0}$ is the value of $n_{e}$ at $t=t_{0}$. Hereafter, we take $v_{w} = 1000$\,km\,s$^{-1}$, the typical value for Galactic Wolf-Rayet stars  \citep[see][]{2012ApJ...752...17W}.

%%%%%%%%%%%%%%%%%%%%%%%%%%%%%%%%%%%%%%%%%%%%%%%%%%%%%%%%%%%%%%%%%%%%%%%%%%%%%%%%%%
\section{Results}\label{sec:results}
\begin{figure*}
\includegraphics{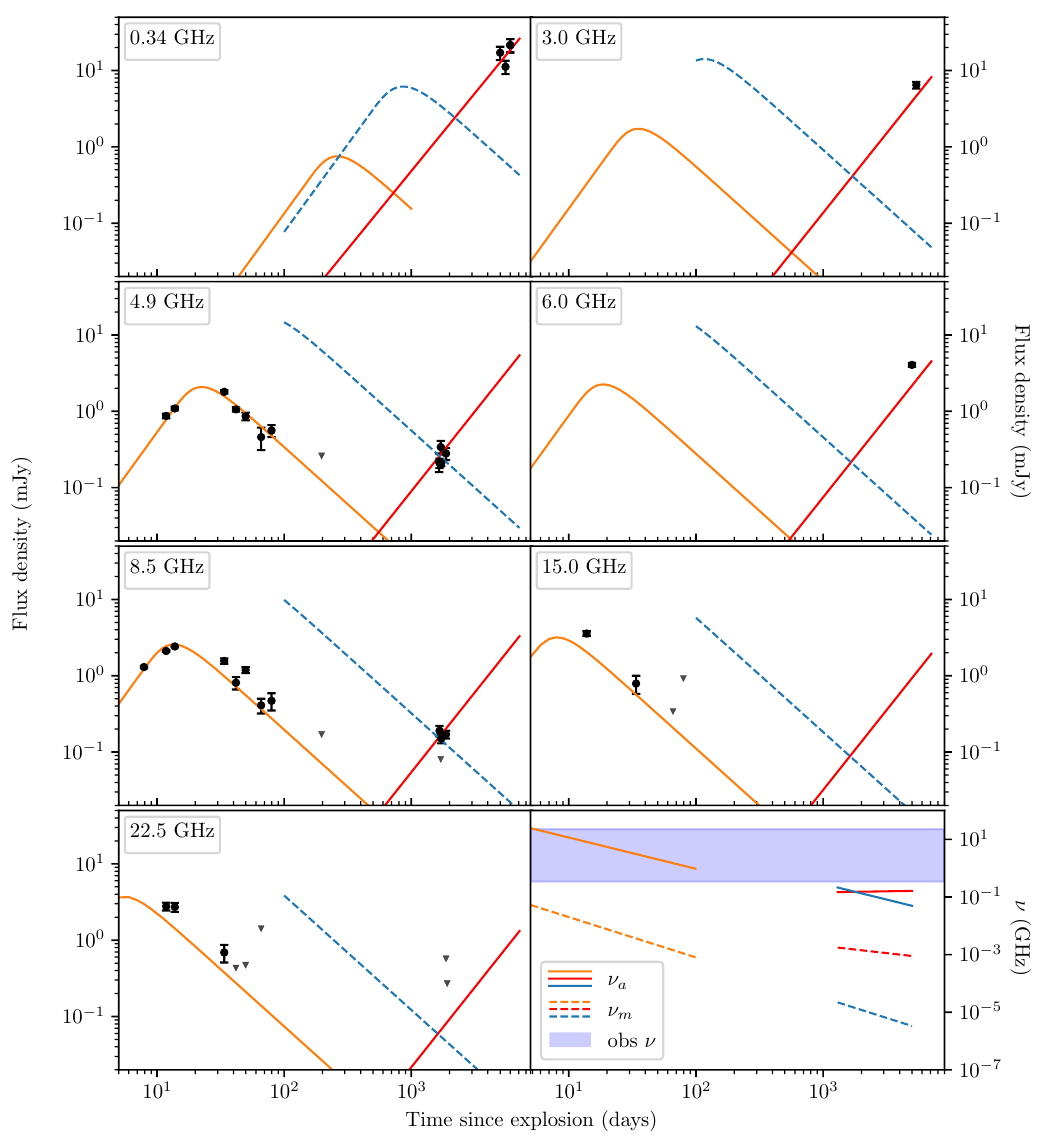}
\caption{The radio light curves of SN\,2004dk (data points) with the best fit models (lines; see also Table \ref{tab:best_fit}). Upper limits are shown as downward pointing triangles and are not included in the fits. The orange curve agrees with the fit obtained by \cite{2012ApJ...752...17W}. Curves at later times (red and blue) are fits derived assuming a smooth radial evolution of the SN ejecta. The bottom right panel of the plot shows the temporal evolution of $\nu_{a}$ (solid lines) and $\nu_{m}$ (dashed lines) and the color code is the same as that of the light curves. The extent of the light blue patch covers the observation frequencies from the lowest 340\,MHz to the highest 22.5\,GHz.}
\label{fig:lc_plots}
\end{figure*}

\begin{figure*}
\centering
\includegraphics[scale=0.8]{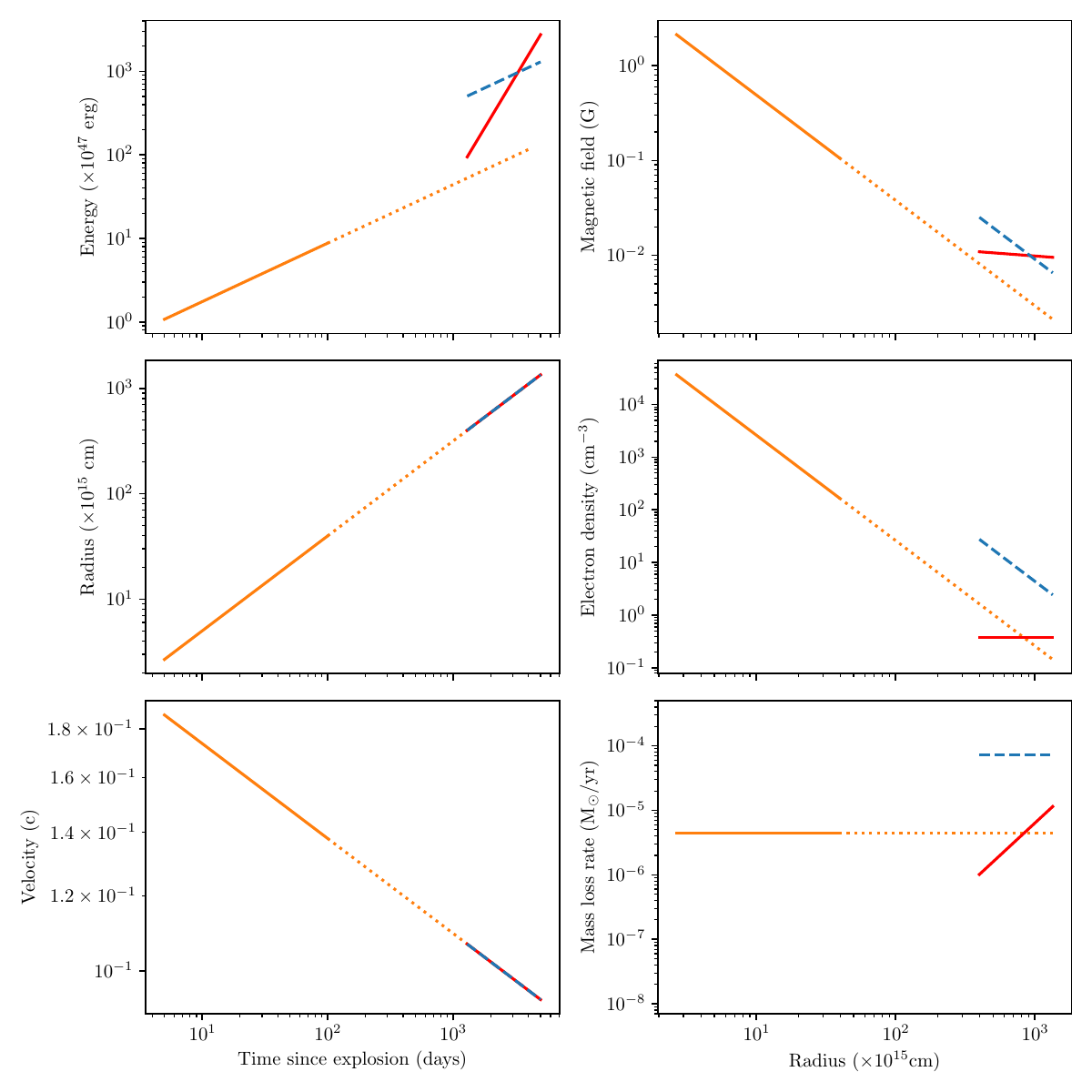}
\caption{Plot of physical parameters for various fits using the same colors as the light curve plot (Figure \ref{fig:lc_plots}). The panels on the left show the temporal evolution of the shocked shell properties, while the panels on the right show the radial evolution of the ambient medium and magnetic field. See text for discussion.\label{fig:phys_param_plots}}
\end{figure*}

Using all detections reported in Table \ref{tab:data}, we perform a $\chi^{2}$ fit of the SN\,2004dk radio observations within the SSA model described in the previous Section. For all fits, we set $t_{0}=10$\,days from the explosion time $t_e$ and, following \cite{2012ApJ...752...17W}, $\eta=4$. Table \ref{tab:best_fit} shows the best fit results obtained for various scenarios. The orange fit (and corresponding orange lines in Figures \ref{fig:lc_plots} and \ref{fig:phys_param_plots}) considers only data obtained within the first $\sim$ 100\,days from the SN\,2004dk explosion, so as to exclude the radio re-brightening phase. For this fit, following the ``standard'' fit by \cite{2012ApJ...752...17W}, we fix $\alpha_{r}=0.9$, $p=3.0$, $s=2.0$, and $\xi=1.0$. We further set $\nu_{m,0}=0.02$\,GHz \citep{2006ApJ...651.1005S} and allow $B_{0}$ and $r_{0}$ to vary. The obtained best fit values for $B_{0}$ and $r_{0}$ agree with those by \cite{2012ApJ...752...17W}. This fit yields a $\chi^{2}/dof \approx  174/18$.
\setlength\LTcapwidth{\linewidth}
\begin{longtable}{ccccc}
\caption{The best fit results for SN\,2004dk within the SSA model. See text for discussion.
\label{tab:best_fit}}\\
\toprule
\toprule
Parmeters & Orange fit & Red fit & Blue fit \\
\midrule
$B_{{0}}$ (G) & 1.06 & 0.02 & 3.27 \\
$\alpha_{{r}}$ & 0.9 $^{f}$ & 0.9 $^{f}$ & 0.9 $^{f}$ \\
$p$ & 3.0 $^{f}$ & 3.0 $^{f}$ & 3.0 $^{f}$ \\
$\nu_{{m,0}}$ (GHz) & 0.02 $^{f}$ & 0.02 $^{f}$ & 0.02 $^{f}$ \\
$r_{{0}}$ ($\times 10^{{15}}$ cm) & 5.0 & 5.0 $^{f}$ & 5.0 $^{f}$ \\
$s$ & 2.0 $^{f}$ & 0.0 & 2.0 $^{f}$ \\
$t_{{e}}$ (JD) & 2453216.7 $^{f}$ & 2453216.7 $^{f}$ & 2453216.7 $^{f}$ \\
$\xi$ & 1.0 $^{f}$ & 0.2 & 0.9 \\
\midrule
$\chi^{2}$/dof & 173.80/18 & 100.98/11 & 495.45/12\\
\midrule
$\gamma_{{m,0}}$ & 2.6 & 20.1 & 1.5 \\
$E_{{0}}$ (erg) & 1.8 $\times 10^{{47}}$ & 4.9 $\times 10^{{43}}$ & 1.7 $\times 10^{{48}}$ \\
$v_0$ (c) & 0.17 & 0.17 & 0.17 \\
$n_{{e,0}}$ (cm$^{{-3}}$) & 1.0 $\times 10^{{4}}$ & 3.8 $\times 10^{{-1}}$ & 1.8 $\times 10^{{5}}$ \\
$\Dot{M}_{{0}}$ (M$_{{\odot}}$/yr) & 4.4 $\times 10^{{-6}}$ & 1.6 $\times 10^{{-10}}$ & 7.4 $\times 10^{{-5}}$ \\
\bottomrule
\multicolumn{4}{l}{$^{f}$ fixed parameter}
\end{longtable}

\begin{figure*}
    \centering
    \includegraphics[width=16cm]{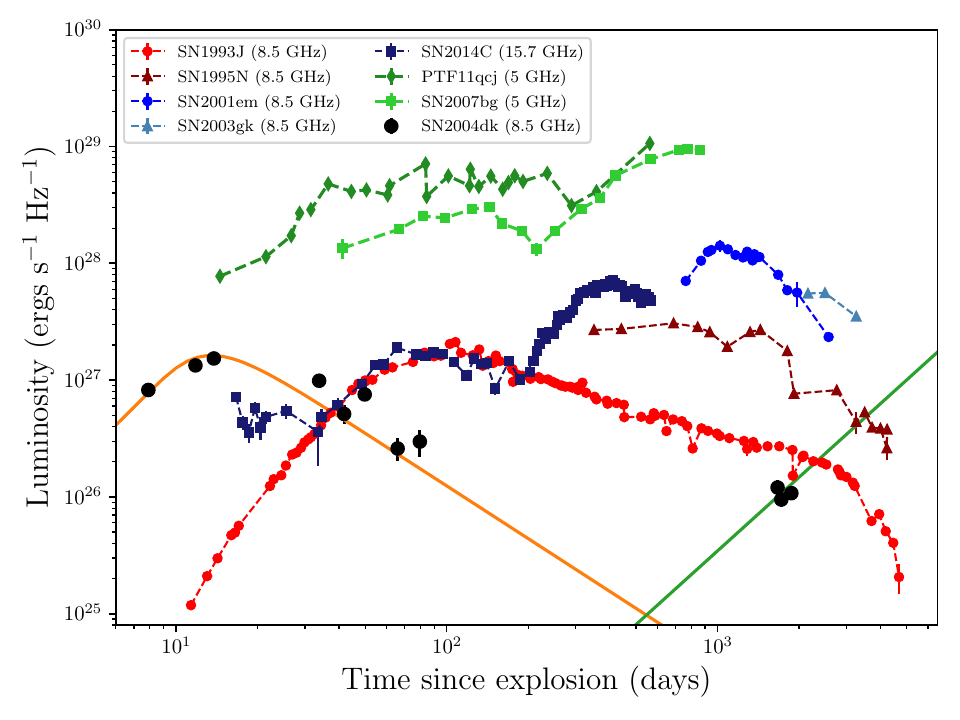}
    \caption{8.5\,GHz lightcurve of SN\,2004dk (data points in black with the best fit lightcurves for both early and late times as plotted in \ref{fig:lc_plots}) compared with late-time lightcurves of other SNe (type II in shades of red, type Ib in shades of blue and type Ic BL in shades of green). Data for SN\,1993J from \cite{2007ApJ...671.1959W}; SN\,1995N from \cite{2009ApJ...699..388C}; SN\,2001em from \cite{2020ApJ...902...55C}; SN\,2003gk from \cite{2014MNRAS.440..821B}; and SN\,2014C from \cite{2017MNRAS.466.3648A}:SN\,2007bg from \cite{2013MNRAS.428.1207S}; and PTF11qcj from \cite{2014ApJ...782...42C} and \cite{2019ApJ...872..201P}.}
    \label{fig:comp_SN}
\end{figure*}

Next, we perform an independent analysis  of the late-time ($t\gtrsim  1000$\,d since explosion) radio observations of SN\,2004dk within the SSA model. All the fits assume a smooth radial evolution of the SN\,2004dk ejecta with $\alpha_r=0.9$ and $r_{0}=5.0\times10^{15}$\,cm (as derived from the fit of the early-time data). For the red fit (and corresponding red lines in Figures \ref{fig:lc_plots} and \ref{fig:phys_param_plots}) we fix $\nu_{m,0}=0.02$\,GHz, and $p=3.0$, and vary $B_{0}$  and $s$ and $\xi$. The best fit result yields $s\approx0$ and a $\chi^{2}/dof\approx  101/11$. 
For comparison, we also report a blue fit (and corresponding blue lines in Figures \ref{fig:lc_plots} and \ref{fig:phys_param_plots}) where we set $s= 2$ (standard case of a smooth CSM shaped by a constant mass-loss rate from the progenitor). We see in Figure \ref{fig:lc_plots} that the blue curve does not fit the data well ($\chi^{2}/dof\approx  495/12$). 

We finally note that the results reported in Table \ref{tab:best_fit} are somewhat dependent on the assumed value of $\eta$, given the scalings $B_0\propto \eta^{4/17}$, $r_0\propto \eta^{1/17}$, and $\gamma_{m,0}\propto \eta^{-2/17}$  \citep[see Equations (5)-(7) in][]{2006ApJ...651.1005S}.
%%%%%%%%%%%%%%%%%%%%%%%%%%%%%%%%%%%%%%%%%%%%%%%%%%%%%%%%%%%%%%%%%%%%%%%%%%%%%%%%%%

\section{Discussion and conclusion} \label{sec:disc}
From the outcome of the fits reported in the previous Section, it is evident that the standard $s=2$ (constant mass-loss rate) and $p=3$ scenario of the SSA model for radio SNe fits the SN\,2004dk radio data well at early times ($t\lesssim 100$\,days since explosion, orange fit in Figure. \ref{fig:lc_plots}) and agrees with the fit obtained by \cite{2012ApJ...752...17W}. However, interpreting the SN\,2004dk late-time radio re-brightening within the SSA scenario requires a modified CSM profile which approaches the $s=0$ case of a constant density CSM (compare the blue fit with the red one). 

Changes in the mass-loss rate profile at late times have been observed in other SNe with peculiar and variable radio light curves \citep[e.g.,][]{1998ApJ...506..874M,2012ApJ...752...17W,2014ApJ...782...42C,2019ApJ...872..201P} and can be attributed to a significant evolution in the mass-loss history of the SN progenitor in the years before explosion, perhaps related to precursor eruptions to the main SN event. In Figure \ref{fig:phys_param_plots}, we show the evolution of the mass-loss rate with radius (bottom right panel) for the late-time best fit models of SN\,2004dk radio light curves, compared to that derived from the early-time data (orange). 

It is particularly interesting that the fit providing the best $\chi^2/dof$ pushes the parameter $s$ to $s=0$, so that a constant ISM density profile seems to be favored over a stellar wind one \citep[see][for a discussion of the $s=0$ case]{Chevalier1982}. To better understand the reasons behind this result, the following considerations are useful. As shown in the bottom-right panel in Figure \ref{fig:lc_plots}, our observing radio frequencies $\nu$ (light blue patch) are such that, for all best fit models considered here, $\nu_m< \nu_{a} < \nu < \nu_{c}$. Thus, from Equation (\ref{eq:f_nu}), we expect the flux to scale as:
\begin{equation}\label{eqn:main_f_nu}
    f_{\nu} \propto t^{\frac{1}{4} (6\alpha_{r} - 5s\alpha_{r} + 10p\alpha_{r} -10p -sp\alpha_{r} +6)}.
\end{equation}
As is evident from the above Equation, when $s=0$ (and for $\alpha_r=0.9$ and $p=3$), $f_{\nu}$ increases with time in this frequency regime, while for $s=2$ (blue fit in Figure \ref{fig:lc_plots}) the flux decrease with time, thus making it hard to fit a re-brightening behavior such as the one observed in SN\,2004dk (see Figure \ref{fig:lc_plots}).

From Equations (\ref{eq:M_dot})-(\ref{eqn:main_f_nu}), when $s\ne 2$ we can further derive the following scaling of the observed flux with the mass-loss rate:
\begin{equation}
    f_{\nu} \propto  \left(\frac{\Dot{M}}{v_{w}}\right)^{\frac{[\alpha_{r}(6 - 5s + 10p -sp) - 10p + 6]}{(4\alpha_{r}(2 - s))}}.
\end{equation}
For $\alpha_r=0.9$, $p=3$. and $s=0$, one gets: 
\begin{equation}
    f_{\nu} \propto  \left(\frac{\Dot{M}}{v_{w}}\right)^{+1.2}
    \vspace{5pt}
\end{equation}
which shows how an increase in mass-loss rate corresponds to a re-brightening of the observed flux (see Figure. \ref{fig:phys_param_plots} for the radial evolution of the mass-loss rate, and Figure \ref{fig:lc_plots} for the temporal evolution of the flux density). 

Now, we can compare our results with those reported in \cite{2018MNRAS.478.5050M}, where the velocity measurements from late-time observations of the H$\alpha$ profiles yielded $v_{w}\approx 400$\,km s$^{-1}$. Assuming this to be the speed of an H-rich CSM outflow from the progenitor system, we find that the H-rich material was ejected $\approx 400$\, years before explosion (since the SN shock meets this CSM outflow at a radius of $r\sim 5\times10^{17}$\,cm. The last is derived using the radius of the shell as predicted by the best fit models shown in red and blue in Figure \ref{fig:phys_param_plots}, at the time of the earliest radio data used in those fits). This is an approximate estimate of the boundary of the H-rich CSM medium. This estimate closely agrees with the $\approx320$\,years found by \cite{2018MNRAS.478.5050M}, which suggests a ``superwind'' related to pulsations of the partially ionised H envelope during the red supergiant phase \citep{1997A&A...327..224H}. Computational studies of such a mechanism show that winds can cause episodic mass loss from the star, all in a timescale of $\lesssim 10^{3}$ years before the explosion \citep{2010ApJ...717L..62Y}. 

\cite{2019ApJ...883..120P} interpret the late-time X-ray and optical behavior of SN\,2004dk in the ``wind bubble model", where the emission is a result of interaction with a CSM prepared by inner fast winds interacting with a previous slow wind. In this framework, the inner CSM density follows the usual radial dependence, $\rho \propto r^{-2}$, agreeing with the SSA fit obtained at early times in this work and in \citet{2012ApJ...752...17W}. However, the shock will then interact with a nearly constant density region before reaching the cold, dense shell (CDS), which is formed due to radiative cooling of matter in the region where the fast and slow winds collide. While a detailed comparison with the model by \citet{2019ApJ...883..120P} is beyond the scope of our analysis, we note that our best fit SSA model at late-times (red curves in Figure \ref{fig:phys_param_plots}) also hints at a constant density medium. 

In Figure \ref{fig:comp_SN} we compare the radio light curve of SN\,2004dk with other radio SNe whose light curves have been monitored on long timescales. The radio light curve of the Type IIb SN\,1993J has been modelled via a combination of free-free absorption and synchrotron self-absorption \citep[see Section \ref{sec:model} and][]{2007ApJ...671.1959W}. The bumpy radio light curve of the Type IIn SN\,1995N has been explained by invoking density enhancements in the CSM \citep{2009ApJ...699..388C}. SN\,2001em, initially classified as a Type Ib explosion, has shown evidence for late-time interaction with an H-rich CSM (ejected by the progenitor a few thousand years before explosion). It has thus been re-classified as a Type IIn SN and also as the second brightest explosion of this type in the radio \citep{2020ApJ...902...55C}. In the case of the Type Ib SN\,2003gk, a decaying late-time radio emission has been observed several  years after explosion  \citep{2014MNRAS.440..821B}. SN\,2014C represents another case of a Type IIn explosion with two distinct phases of mass loss invoked for its progenitor, on timescales similar to that of SN\,2004dk \citep{2017MNRAS.466.3648A}. PTF11qcj and SN\,2007bg are the most radio-loud Type Ic with broad lines, showing evidence for interaction with a CSM of varying density \citep[][]{2014ApJ...782...42C,2019ApJ...872..201P,2013MNRAS.428.1207S}, again qualitatively similar to the case of SN\,2004dk. We note that the 8.5\,GHz radio peak luminosity of SN\,2004dk during its first peak of emission is comparable to the peak luminosities of the first peak of SN\,2014C and SN\,1993J (Figure \ref{fig:comp_SN}).

In conclusion, the VLITE observations reported in this paper, combined with previously collected VLA data,  favor  an interpretation of SN\,2004dk as a  strongly  CSM-interacting  radio SN going trough an environment shaped by the progenitor wind at first, and then followed by an ISM-like profile. The re-brightening associated with this change in density profile was first probed in 2009 \citep[see][]{2009CBET.1714....1S,2012ApJ...752...17W,2019ApJ...883..120P}. Our VLITE observations confirm such re-brightening episode and suggest that the radio emission of SN\,2004dk will likely continue to brighten at GHz frequencies until a second peak in the light curve is reached. However, our conclusions are also limited by the simplifications inherent in the spherically symmetric SSA model we adopted.  Continued VLA/VLITE observations will be able to further test this hypothesis (or else spur a more complex theoretical modeling of this peculiar radio SN light curve).

\acknowledgements
 A.B. and A.C. acknowledges support from the National Science Foundation CAREER  Award  \#1455090. The National Radio Astronomy Observatory is a facility of the National Science Foundation operated under cooperative agreement by Associated Universities, Inc. Basic research in radio astronomy at the Naval Research Laboratory is funded through 6.1 Base funding.

\bibliography{references}
\end{document}